\documentstyle[12pt,aasms4]{article}
\def\be{\begin{equation}}
\def\ee{\end{equation}}
\def\ba{\begin{eqnarray}}
\def\ea{\end{eqnarray}}

\def\fun#1#2{\lower3.6pt\vbox{\baselineskip0pt\lineskip.9pt
        \ialign{$\mathsurround=0pt#1\hfill##\hfil$\crcr#2\crcr\sim\crcr}}}

\slugcomment{To be submitted to ApJ}

\begin{document}
\null\vspace{-62pt}
\begin{flushright}
astro-ph/9901212\\
June 25, 1999
\end{flushright}
\title{Analytical Modeling of the Weak Lensing of Standard Candles\\
I. Empirical Fitting of Numerical Simulation Results}

\author{Yun Wang}
\affil{{\it Princeton University Observatory} \\
{\it Peyton Hall, Princeton, NJ 08544\\}
{\it email: ywang@astro.princeton.edu}}

\vspace{.4in}
\centerline{\bf Abstract}
\begin{quotation}

Weak lensing leads to the non-Gaussian magnification distribution 
of standard candles at given redshift $z$, $p(\mu|z)$. 
In this paper, we give accurate and simple empirical fitting 
formulae of the weak lensing numerical simulation results with 
the generalized Dyer-Roeder prescription.
The smoothness parameter $\tilde{\alpha}$ essentially represents
the amount of matter that can cause magnification of a given source.
Since matter distribution in our universe is inhomogeneous, we can think
of our universe as a mosaic of cones centered on the observer, 
each with a different value of $\tilde{\alpha}$.
We define the {\it  direction dependent} smoothness parameter $\tilde{\alpha}$
via the Dyer-Roeder equation; there is a unique mapping between 
$\tilde{\alpha}$ and the magnification of a source.
We find that the distribution of $\tilde{\alpha}$ at given $z$, 
$p(\tilde{\alpha}|z)$, is well described by a modified Gaussian distribution.
For the same matter distribution, i.e., the same $p(\tilde{\alpha}|z)$,
different values of $\Omega_m$ and $\Omega_{\Lambda}$ can lead to very
different magnification distributions.

Our formulae can be conveniently used to calculate the weak lensing
effects for observed Type Ia supernovae at arbitrary redshifts.

\end{quotation}


\section{Introduction}

The use of standard candles is fundamental in observational cosmology.
The distance-redshift relations for standard candles enable us to 
determine the basic cosmological parameters $H_0$ (the Hubble constant),
$\Omega_m$ (the matter density of the Universe in units of the 
critical density $\rho_c=3H_0^2/(8\pi G)$), and $\Omega_{\Lambda}$
(the density contribution from the cosmological constant in units
of $\rho_c$). 

At present, the best candidates for standard candles are Type Ia supernovae
(SNe Ia), because they can be calibrated to have very small intrinsic 
dispersions at cosmological distances (\cite{Phillips93,Riess95}).
Two independent groups of observers (\cite{Perl98,Riess98}) have
demonstrated that SNe Ia can be potentially powerful tools for cosmology.
Their current results (\cite{Perl99,Schmidt98})
seem to indicate a low matter density universe, possibly with a 
sizable cosmological constant. It is clear that 
the observation of SNe Ia can potentially become a reliable 
probe of cosmology. 
However, there are important systematic uncertainties of SNe Ia as 
standard candles, in particular, evolution and gravitational lensing.
The two groups have either assumed a smooth universe
in their data analysis, or included
lensing effects in rudimentary ways. 
Since we live in a clumpy universe,
the effect of gravitational lensing must be taken into account
adequately for the proper interpretation of SN data.

A number of authors have considered various aspects of
the gravitational lensing of SNe Ia 
(\cite{Frieman97,Wamb97,Kantow98,Holz98b,Holz98,Metcalf98,Porciani98}).
A realistic calculation of the weak lensing effect of standard
candles was conducted by Wambsganss et al. (1997), who computed 
the magnification distributions of standard candles
using a N-body simulation which has a resolution on small
scales that is of the order the size of a halo.
However, since the magnification distributions depend on the
cosmological model and redshift, the numerical results 
can not be directly used to 
compute the effect of weak lensing for an observed SN Ia
at arbitrary redshift.

In this paper we derive accurate empirical fitting formulae 
of the weak lensing numerical simulation results for the distribution 
of the magnification of standard candles due to weak lensing; these 
simple formulae can be used to account for the effect of weak lensing 
in the analysis of SN Ia data.
In \S 2, we give analytical formulae for the angular diameter distance 
of a standard candle in terms of the smoothness parameter $\tilde{\alpha}$ 
and constants which depend on redshift and the cosmological parameters 
$\Omega_m$ and $\Omega_{\Lambda}$.
In \S 3, we generalize the interpretation of the angular diameter 
distance obtained in \S 2 by allowing the smoothness parameter 
$\tilde{\alpha}$ to be a direction dependent variable,
the direction dependent smoothness parameter; 
we extract the distributions of $\tilde{\alpha}$
from the magnification distributions found by numerical simulations.
In \S 4, we give analytical formulae for computing the magnification
distributions at arbitrary redshifts. \S 5 contains a summary and discussions.

\section{Angular diameter distance as function of the smoothness 
parameter $\tilde{\alpha}$}

In a Hubble diagram of standard candles, one must use
distance-redshift relations to make theoretical interpretations. 
The distance-redshift relations depend on the distribution of
matter in the universe. 
In this section, we express the angular diameter
distance to a standard candle in terms of the smoothness parameter,
$\tilde{\alpha}$, which is 
the mass-fraction of the matter in the universe smoothly distributed
(\cite{DR73}).

In a smooth Friedmann-Robertson-Walker (FRW) universe,
$\tilde{\alpha}=1$ in all beams;
the metric is given by $ds^2=dt^2-a^2(t)[dr^2/(1-kr^2)+r^2 (d\theta^2
+\sin^2\theta \,d\phi^2)]$, where $a(t)$ is the cosmic scale factor,
and $k$ is the global curvature parameter ($\Omega_k
=1-\Omega_m-\Omega_{\Lambda}=-k/H_0^2$).
The comoving distance $r$ is given by (\cite{Weinberg72})
\be
\label{eq:r(z)}
r(z)=\frac{cH_0^{-1}}{|\Omega_k|^{1/2}}\,
\mbox{sinn}\left\{ |\Omega_k|^{1/2}
\int_0^z dz'\,\left[ \Omega_m(1+z')^3+\Omega_{\Lambda}+\Omega_k(1+z')^2
\right]^{-1/2} \right\},
\ee
where ``sinn'' is defined as sinh if $\Omega_k>0$, and sin if  $\Omega_k<0$.
If $\Omega_k=0$, the sinn and $\Omega_k$'s disappear from Eq.(\ref{eq:r(z)}),
leaving only the integral. 
The angular diameter distance is given by $d_A(z)=r(z)/(1+z)$,
and the luminosity distance is given by $d_L(z)=(1+z)^2 d_A(z)$.

However, our universe is clumpy rather than smooth.
According to the focusing theorem in gravitational lens theory,
if there is any shear or matter along a beam connecting 
a source to an observer, the angular diameter distance of
the source from the observer is {\it smaller} than that which would occur
if the source were seen through an empty, shear-free cone,
provided the affine parameter distance (defined such that its element
equals the proper distance element at the observer) 
is the same and the beam has not gone through a caustic.
An increase of shear or matter density along the beam decreases the
angular diameter distance and consequently increases the
observable flux for given $z$. (Schneider, Ehlers, \& Falco 1992)

For given redshift $z$,
if a mass-fraction $\tilde{\alpha}$ of the matter in the universe is smoothly
distributed, the largest possible distance for
light bundles which have not passed through a caustic is given by
the solution to the following equation:
\ba
\label{eq:DR}
&& g(z) \, \frac{d\,}{dz}\left[g(z) \frac{dD_A}{dz}\right]
+\frac{3}{2} \tilde{\alpha} \,\Omega_m (1+z)^5 D_A=0, \nonumber \\
&& D_A(z=0)=0, \hskip 1cm \left.\frac{dD_A}{dz}\right|_{z=0}=\frac{c}{H_0},
\ea
where $g(z) \equiv (1+z)^3 \sqrt{ 1+ \Omega_m z+ \Omega_{\Lambda}
[(1+z)^{-2} -1] }$ (\cite{Kantow98}).
The $\Omega_{\Lambda}=0$ form of Eq.(\ref{eq:DR}) has been known as
the Dyer-Roeder equation (\cite{DR73,Sch92}).

The angular diameter distance for given smoothness parameter $\tilde{\alpha}$
and redshift $z$, $D_A(\tilde{\alpha},z)$, can be obtained via the proper 
integration of Eq.(2). However, for our purposes,
it is useful to fit the angular diameter distance at given redshift,
$D_A(\tilde{\alpha}|z)$, to a polynomial in $\tilde{\alpha}$, with the 
coefficients dependent on $z$ and cosmological parameters. 
For the ranges of $\tilde{\alpha}$ and $z$ of interest, the angular diameter
distance given by Eq.(\ref{eq:DR}) can be approximated as
\be
\label{eq:DRa}
D_A(\tilde{\alpha}|z)\simeq cH_0^{-1}\left[ d_0(z)+ a(z) 
\tilde{\alpha}^3+b(z) \tilde{\alpha}^2 + c(z)\tilde{\alpha}\right],
\ee
where $d_0=D_A(\tilde{\alpha}=0)$, and
\ba
\label{eq:abc}
a(z) &= & \frac{4}{3} \left(-d_0+3d_{0.5}
-3d_1+d_{1.5} \right), \nonumber\\
b(z) &=& 2 \left( d_0-2d_{0.5}+d_1-\frac{3a}{4} \right), \nonumber\\
c(z) &=& -d_0+d_1-a-b,
\ea
with $d_{0.5}=D_A(\tilde{\alpha}=0.5)$, $d_{1}=D_A(\tilde{\alpha}=1)$, and
$d_{1.5}=D_A(\tilde{\alpha}=1.5)$.
Fig.1 shows that the difference between Eq.(\ref{eq:DRa}) and
the solution to Eq.(\ref{eq:DR}) is much smaller than 0.1 \% for the ranges
of interest (see Fig.3).

The constants $d_0$, $d_{0.5}$, $d_{1}$, and $d_{1.5}$ depend on
the redshift $z$ and the cosmological parameters $\Omega_m$ and 
$\Omega_{\Lambda}$; they are easily computed by integrating Eq.(\ref{eq:DR}).
Table 1 lists $d_0$, $d_{0.5}$, $d_{1}$, and $d_{1.5}$
for various cosmological models at various redshifts.

\begin{table}[htb]
\caption{The constants $d_0$, $d_{0.5}$, $d_{1}$, and $d_{1.5}$
from Eqs.(\ref{eq:DRa}) and (\ref{eq:abc}), for various cosmological
models and redshifts.}
\begin{center}
\begin{tabular}{ccccccccc}
\hline\hline
($\Omega_m$,$\Omega_{\Lambda}$) & &  $z=0.5$ & $z=1.0$ & $z=1.5$ & $z=2.0$ 
& $z=2.5$ & $z=3.0$ & $z=5.0$ \\ 
\hline
(0.4, 0.6) & 
$d_0$ & 0.29088 & 0.39373 & 0.43849 & 0.46109 & 0.47378 & 0.48150 & 0.49400 \\
&$d_{0.5}$& 0.28766 & 0.37929 & 0.40897 & 0.41577 & 0.41329 & 0.40692 &0.37412\\
&$d_1$ &0.28445 &0.36517 & 0.38069 & 0.37329 & 0.35782 & 0.34000 & 0.27469 \\
&$d_{1.5}$&0.28127& 0.35138 &0.35362 &0.33352 &0.30708&0.28015 &0.19307\\
\hline
(0.2, 0.8) &
$d_0$ & 0.30893 & 0.43237 & 0.49007 & 0.52031 & 0.53764 & 0.54831 & 0.56578 \\
&$d_{0.5}$ & 0.30694 & 0.42198& 0.46655& 0.48168& 0.48361 & 0.47940 & 0.44655 \\
&$d_1$ & 0.30496 & 0.41175 & 0.44374 & 0.44485 & 0.43304 & 0.41610 & 0.34456 \\
&$d_{1.5}$& 0.30298 & 0.40166 &0.42160 & 0.40976 & 0.38577 & 0.35808& 0.25795\\
\hline
(0.2, 0) &
$d_0$ & 0.27246 & 0.36342 & 0.40377 & 0.42485 & 0.43710 & 0.44480 & 0.45790\\
&$d_{0.5}$ &0.27116&0.35793 & 0.39252 &0.40722& 0.41294 & 0.41415 & 0.40342\\
&$d_1$ & 0.26987 & 0.35248 & 0.38147 & 0.39006& 0.38963 & 0.38488 & 0.35335\\
&$d_{1.5}$&0.26858& 0.34709 & 0.37062 & 0.37336 & 0.36716 &0.35693& 0.30743\\
\hline
(1, 0) &
$d_0$ & 0.25485 & 0.32929 & 0.35952 & 0.37434 & 0.38255 & 0.38750 & 0.39546 \\
&$d_{0.5}$&0.24973& 0.31077& 0.32577& 0.32602 & 0.32096 & 0.31401 & 0.28527 \\
&$d_1$ & 0.24467 & 0.29289 & 0.29404& 0.28177 & 0.26599 & 0.25000 & 0.19725 \\
&$d_{1.5}$&0.23968& 0.27565& 0.26424& 0.24134 & 0.21710 & 0.19455 & 0.12797 \\
\hline
\end{tabular}
\end{center}
\end{table}

Note that Eq.(\ref{eq:DR}) is usually used to describe a universe 
with a global smoothness parameter $\tilde{\alpha}$, 
with $0\leq \tilde{\alpha}\leq 1$, since $\tilde{\alpha}$
is the {\it fraction} of matter in the universe which is smoothly
distributed. However, Eq.(\ref{eq:DR}) is well defined for all positive
values of $\tilde{\alpha}$. Since $\tilde{\alpha}$ essentially represents
the amount of matter that causes weak lensing of a given source,
and matter distribution in our universe is inhomogeneous, we can think
of our universe as a mosaic of cones centered on the observer, 
each with a different value of $\tilde{\alpha}$. 
This reinterpretation of $\tilde{\alpha}$ implies
that we have $\tilde{\alpha}>1$ in regions of the universe in which
there are above average amounts of matter which can cause magnification
of a source (see \S 3).

\section{The distribution of the direction dependent 
smoothness parameter $\tilde{\alpha}$}

In the previous section, we showed that we can separate the 
dependence of the angular diameter distance on the mass-fraction 
of matter smoothly distributed ($\tilde{\alpha}$) from 
its dependence on the redshift and cosmological parameters (see Eq.(3)).
Unlike angular separations and flux densities, distances are {\it not} 
directly measurable. 
Let us generalize the angular diameter distance
obtained in \S 2 by allowing the smoothness parameter $\tilde{\alpha}$
to be {\it  direction  dependent}, i.e., a property of the {\it beam} 
connecting the observer and the standard candle.
In order to derive a unique mapping between the distribution
in distances and the distribution in the direction dependent smoothness parameter 
for given redshift $z$, we {\it define} the direction dependent smoothness parameter
$\tilde{\alpha}$ to be the solution of Eq.(\ref{eq:DR}) (or Eq.(3)) 
for given distance $D_A(z)$.

Note that in numerical simulations, as in the real universe,
we can have two lines of sight with the same fraction of smoothly
distributed matter, but different distances to a given redshift $z$,
because weak lensing depends on {\it where} the matter is, as well
as what fraction of the matter is smoothly distributed.
Our definition of the direction dependent smoothness parameter $\tilde{\alpha}$
implies that it is no longer simply the fraction of matter smoothly 
distributed, it also contains information on where matter is
distributed. We can interpretate $\tilde{\alpha}$ as 
the ratio of the {\it effective} density of matter smoothly distributed 
in the beam connecting the observer and the standard candle
and the average matter density in the universe, with the 
effective matter density corresponding to a given amount of magnification.
Two lines of sight with the same fraction of smoothly distributed 
matter but different distances would have different effective 
densities of smoothly distributed matter, thus different values of
$\tilde{\alpha}$.

For given redshift $z$, we expect a {\it distribution} in the
angular diameter distance $D_A(z)$ because the distribution of matter
between redshift zero and redshift $z$ is {\it inhomogeneous}.
We have parametrized matter distribution with $\tilde{\alpha}$,
the direction dependent smoothness parameter;
Eq.(3) then tells us how the angular diameter distance depends
on the matter distribution for given redshift $z$.

Since the direction dependent smoothness parameter $\tilde{\alpha}$ 
describes the distribution of matter in an arbitrary
beam, it is a random variable for given redshift.
The direction dependent smoothness parameter $\tilde{\alpha}$ depends on the 
matter density in the beam connecting the observer and the source,
as well as how the matter is distributed in the beam; for 
matter smoothly distributed throughout the beam,
$\tilde{\alpha}<1$ in underdense beams, while 
$\tilde{\alpha}>1$ in overdense beams.
Note that here we do not consider the possibility that a significant
fraction of matter can be in point masses in some beams.

The matter density field is Gaussian on large scales and non-Gaussian on
small scales, thus we parametrize the probability distribution of
$\tilde{\alpha}$ in a form resembling the Gaussian distribution
(see Eq.(5)).

Wambsganss et al. have found numerically the distributions of the 
magnifications of standard candles at various redshifts, $p(\mu|z)$, 
with $z=$0.5, 1, 1.5, 2, 2.5, 3, 5,
for $\Omega_m=0.4$, $\Omega_{\Lambda}=0.6$ (\cite{Wamb97,Wamb99}).
To extract the distribution of $\tilde{\alpha}$, we use
$\mu = \left[ D_A(\tilde{\alpha}=1)/D_A(\tilde{\alpha}) \right]^2$
(see Eq.(\ref{eq:mu})), where $D_A(\tilde{\alpha})$ is given by Eq.(2) or 
Eq.(3). We find that the distribution of $\tilde{\alpha}$ at given redshift 
$z$ is well described by
\be
\label{eq:p(alpha)}
p(\tilde{\alpha}|z)=C_{norm}\, \exp\left[ -\left( \frac{\tilde{\alpha}-
\tilde{\alpha}_{peak}}
{w \,\tilde{\alpha}^q} \right)^2 \right],
\ee
where $C_{norm}$, $\tilde{\alpha}_{peak}$, $w$, and $q$ depend on $z$ 
and are independent of $\tilde{\alpha}$. Fig.2 shows $C_{norm}$, 
$\tilde{\alpha}_{peak}$, $w$, and $q$ as functions
of $z$; the points are extracted from the numerical $p(\mu|z)$, 
the solid curves are analytical
fits given by
\ba
\label{eq:aq}
C_{norm}(z) &=&   10^{-2} \left[   0.53239
  + 2.79165 \, \left(\frac{z}{5}\right)
  - 2.42315\, \left(\frac{z}{5}\right)^2
  + 1.13844\,\left(\frac{z}{5}\right)^3 \right], \nonumber \\
\tilde{\alpha}_{peak}(z) &=&     1.01350    
   -1.07857  \, \left(\frac{1}{5 z}  \right)
   +2.05019  \, \left(\frac{1}{5z}\right)^2  
   -2.14520  \, \left(\frac{1}{5z}\right)^3, \nonumber  \\
w(z) &=&    0.06375
   + 1.75355  \, \left(\frac{1}{5 z} \right) 
   - 4.99383    \, \left(\frac{1}{5z}\right)^2
   + 5.95852    \, \left(\frac{1}{5z}\right)^3, \nonumber \\
q(z) &=&    0.75045    
   +1.85924 \, \left(\frac{z}{5}   \right)
   -2.91830 \, \left(\frac{z}{5}\right)^2   
   +1.59266 \,\left(\frac{z}{5}\right)^3.
\ea   
$C_{norm}(z)$ is the normalization constant for given $z$.
The parameter $\tilde{\alpha}_{peak}(z)$ indicates the average 
smoothness of the universe at redshift $z$, it increases with $z$
and approaches $\tilde{\alpha}_{peak}(z)=1$ at $z=5$;
the parameter $w(z)$ indicates the width of the distribution in
the direction dependent smoothness parameter $\tilde{\alpha}$,
it decreases with $z$.
The $z$ dependences of $\tilde{\alpha}_{peak}(z)$ and $w(z)$ 
are as expected because as we look back to earlier times, 
lines of sight become more filled in with matter, and
the universe becomes smoother on the average.
The parameter $q(z)$ indicates the deviation of $p(\tilde{\alpha}|z)$
from Gaussianity (which corresponds to $q=0$).

Fig.3 shows $p(\tilde{\alpha}|z)$ for $z=0.5$, 2, and 5. The solid line
is derived from $p(\mu|z)$ found numerically by Wambsganss et al. (1997);
the dotted line is given by Eq.(\ref{eq:p(alpha)}), with coefficients
from Eq.(\ref{eq:aq}); the dot-dash line shows the difference between
the solid curve and the dotted curve;
the dashed line shows the Gaussian distribution
related to Eq.(\ref{eq:p(alpha)}),
\be
\label{eq:pGauss}
p^G(\tilde{\alpha}|z)=C_{norm}\, \exp\left[ -\left( \frac{\tilde{\alpha}-
\tilde{\alpha}_{peak}}
{w \,\tilde{\alpha}_{peak}^q} \right)^2 \right].
\ee
Note that it is difficult to see the dotted lines (our empirical fitting
formulae), because they are so close to the solid lines (the numerical results).
This is not surprising since we have a 4-parameter fit to a
smooth bell-like curve.
Comparisons with Eq.(\ref{eq:pGauss}) show how much $p(\tilde{\alpha}|z)$
deviates from the Gaussian distribution.

For the same matter distribution, i.e., the same $p(\tilde{\alpha}|z)$, 
different values of $\Omega_m$ and $\Omega_{\Lambda}$ can lead to very
different magnification distributions (see \S 4).

\section{The magnification distribution due to weak lensing}

At given redshift $z$,
the magnification of a source can be expressed in terms of the
apparent brightness of the source ${\cal L}(\tilde{\alpha})$, or 
in terms of the angular diameter distance to the source $D_A(\tilde{\alpha})$:
\be
\label{eq:mu}
\mu = \frac{ {\cal L} (\tilde{\alpha})}{ {\cal L} (\tilde{\alpha}=1)}
= \left[ \frac{D_A(\tilde{\alpha}=1)}{D_A(\tilde{\alpha})} \right]^2,
\ee
where ${\cal L} (\tilde{\alpha}=1)$ and $D_A(\tilde{\alpha}=1)$
are the flux of the source and angular diameter distance 
to the source in a completely smooth universe, and $\tilde{\alpha}$
is the direction dependent smoothness parameter (see Eq.(3) and \S 3).
Since distances are not directly measurable, we should interpret 
Eq.(\ref{eq:mu}) as defining a unique mapping between the magnification
of a standard candle at redshift $z$ and the direction dependent smoothness parameter
$\tilde{\alpha}$ at $z$; $\tilde{\alpha}$ parametrizes the direction dependent matter 
distribution in a well-defined manner.

The distribution in the magnification of standard candles placed
at redshift $z$ is
\ba
\label{eq:p(mu)}
p(\mu|z) &= & p(\tilde{\alpha}|z)\, \frac{d\tilde{\alpha}}{d\mu}
= p(\tilde{\alpha}|z)\, \frac{D_A(\tilde{\alpha}=1)}{2 \mu^{3/2}}\,
\left( \frac{\partial D_A}{\partial \tilde{\alpha}} \right)^{-1}
\nonumber \\
&\simeq& p(\tilde{\alpha}|z)\, \frac{D_A(\tilde{\alpha}=1)}{2 \mu^{3/2}}\,
\frac{1}{3a \tilde{\alpha}^2+ 2 b\tilde{\alpha}+c},
\ea
where the parameters $a$, $b$, and $c$ are given by Eq.(\ref{eq:abc});
they depend on $\Omega_m$, $\Omega_{\Lambda}$, and $z$.

Fig.4 shows $p(\mu|z)$ at $z=0.5$, 2, and 5 for the cosmological model
$\Omega_m=0.4$ and $\Omega_{\Lambda}=0.6$. The solid line
is the $p(\mu|z)$ found numerically by Wambsganss et al. (1997);
the dotted line is given by Eq.(\ref{eq:p(mu)}), with $p(\tilde{\alpha}|z)$
given by Eq.(\ref{eq:p(alpha)}). Note the excellent agreement between
our empirical fitting formulae and the numerical results.

Fig.5 shows $p(\mu|z)$ for three cosmological models
at three different redshifts: (a)
$z=0.5$, (b) $z=2$, and (c) $z=5$. 
The three cosmological models are:
(1) $\Omega_m=1$ and $\Omega_{\Lambda}=0$;
(2) $\Omega_m=0.2$ and $\Omega_{\Lambda}=0$;
(3) and $\Omega_m=0.2$ and $\Omega_{\Lambda}=0.8$.
We have computed $p(\mu|z)$ using Eq.(\ref{eq:p(mu)}), 
with $p(\tilde{\alpha}|z)$
given by Eq.(\ref{eq:p(alpha)}).
Note that Fig.5 and Fig.4 have the same matter distribution
(the same $p(\tilde{\alpha}|z)$), but different cosmological parameters.

Models with different cosmological parameters should lead to
somewhat different matter distributions $p(\tilde{\alpha}|z)$.
It would be interesting to compare numerical predictions 
for $p(\tilde{\alpha}|z)$ from N-body simulations for different 
cosmological models. In the context of weak lensing of 
standard candles, we expect the cosmological parameter dependence to
enter primarily through the magnification $\mu$ to direction dependent smoothness 
parameter $\tilde{\alpha}$ mapping at given $z$ (the same $\tilde{\alpha}$ 
corresponds to very different $\mu$ in different cosmologies).

\section{Summary and discussions}

We have derived accurate and simple empirical fitting formulae
to the weak lensing numerical simulation results of Wambsganss et al. (1997).
These empirical formulae can be conveniently used
to compute the weak lensing effect of standard candles
for various cosmological models.
Our formulation is based on the unique mapping between the magnification
of a source and the direction dependent smoothness parameter $\tilde{\alpha}$;
$\tilde{\alpha}$ is the ratio of the effective density of matter 
smoothly distributed in the beam connecting the observer and the source 
and the average density of the universe, with the effective matter 
density corresponding to a given amount of magnification.
We find that the distribution of $\tilde{\alpha}$ is well described
by a modified Gaussian distribution; this is interesting since
the matter density field is Gaussian on large scale and non-Gaussian 
on small scales.

We have derived empirical fitting formulae
for $p(\tilde{\alpha}|z)$ (see \S 3) from the numerical 
magnification distributions, $p(\mu|z)$, found by Wambganss et al. (1997)
for $\Omega_m=0.4$, $\Omega_{\Lambda}=0.6$.
For the same matter distribution, i.e., the same $p(\tilde{\alpha}|z)$,
different values of $\Omega_m$ and $\Omega_{\Lambda}$ can lead to very
different magnification distributions (see \S 4).
It would be interesting to see how $p(\tilde{\alpha}|z)$
depends on the cosmological model (\cite{Wang99}).

Our empirical formulae can be used to calculate the weak lensing
effects for observed Type Ia supernovae in general cosmologies
and at arbitrary redshifts. 
At redshifts of a few, the dispersion in SN Ia luminosities due to
weak lensing will become comparable or exceed the intrinsic dispersion
of SNe Ia (\cite{Wang98}). 
The Next Generation Space Telescope (NGST) can detect SNe Ia at as 
high redshifts as they possibly exist; while there are theoretical
uncertainties on the estimated SN Ia rate at high $z$, it is likely that
the NGST will see quite a few SNe Ia at redshifts of a few (\cite{Stockman98}).
Complementing the NGST search, possible
new ground based supernova pencil beam surveys can yield between
dozens to hundreds of SNe Ia per 0.1 redshift interval up to at least
$z=1.5$ (\cite{Wang98}). The systematic uncertainties of SNe Ia as
standard candles will likely be well understood within the next decade.

We have used the numerical results by Wambsganss et al. (1997)
in deriving our analytical formulae, which accurately describe
the non-Gaussian magnification distributions of standard candles
due to weak lensing. We note that
Frieman (1997) gave analytical estimates of the magnification 
dispersions due to weak lensing which are of the same order of 
magnitude as the numerical results of Wambsganss et al. (1997), 
even though he did not consider the non-Gaussian nature of
the magnification distribution.
It would be useful to have numerical results for cosmologies
other than the $\Omega_m=0.4$, $\Omega_{\Lambda}$=0.6 model studied
by Wambsganss et al. (1997).
Holz \& Wald (1998) studied a few cosmological models,
including $\Omega_m=1$, $\Omega_{\Lambda}$=0; they computed their
magnification distributions assuming that the mass in the universe
is distributed in unclustered isothermal spheres.
It would be interesting to compare how different assumptions affect
the numerical results. 

Note that we have used the magnification distributions calculated by
Wambsganss et al. (1997) using a N-body simulation which has a 
resolution on small scales that is of the order the size of a halo.
These magnification distributions contain the information on how
matter is distributed, including the clustering of galaxies.
By generalizing the smoothness parameter $\tilde{\alpha}$ to a 
direction dependent variable, we have been able to describe the weak 
lensing of standard candles in a simple manner, leading to accurate empirical
formulae which can be easily used to calculate the weak lensing effect
of type Ia supernovae. The distributions of $\tilde{\alpha}$ also
contain the information on how matter is distributed.
The derivation of the distribution of $\tilde{\alpha}$ 
from the matter power spectrum should reveal how the measurement 
of $p(\tilde{\alpha}|z)$ (via $p(\mu|z)$) can probe the clustering of matter 
and structure formation in the universe (\cite{Wang99}).

\acknowledgements{\centerline{\bf Acknowledgements}}

It is a pleasure for me to thank Joachim Wambsganss for generously providing 
the magnification distributions used to fit $p(\tilde{\alpha}|z)$,
and for helpful suggestions; and Liliya Williams for very useful comments.


\clearpage
\setcounter{figure}{0}

\figcaption[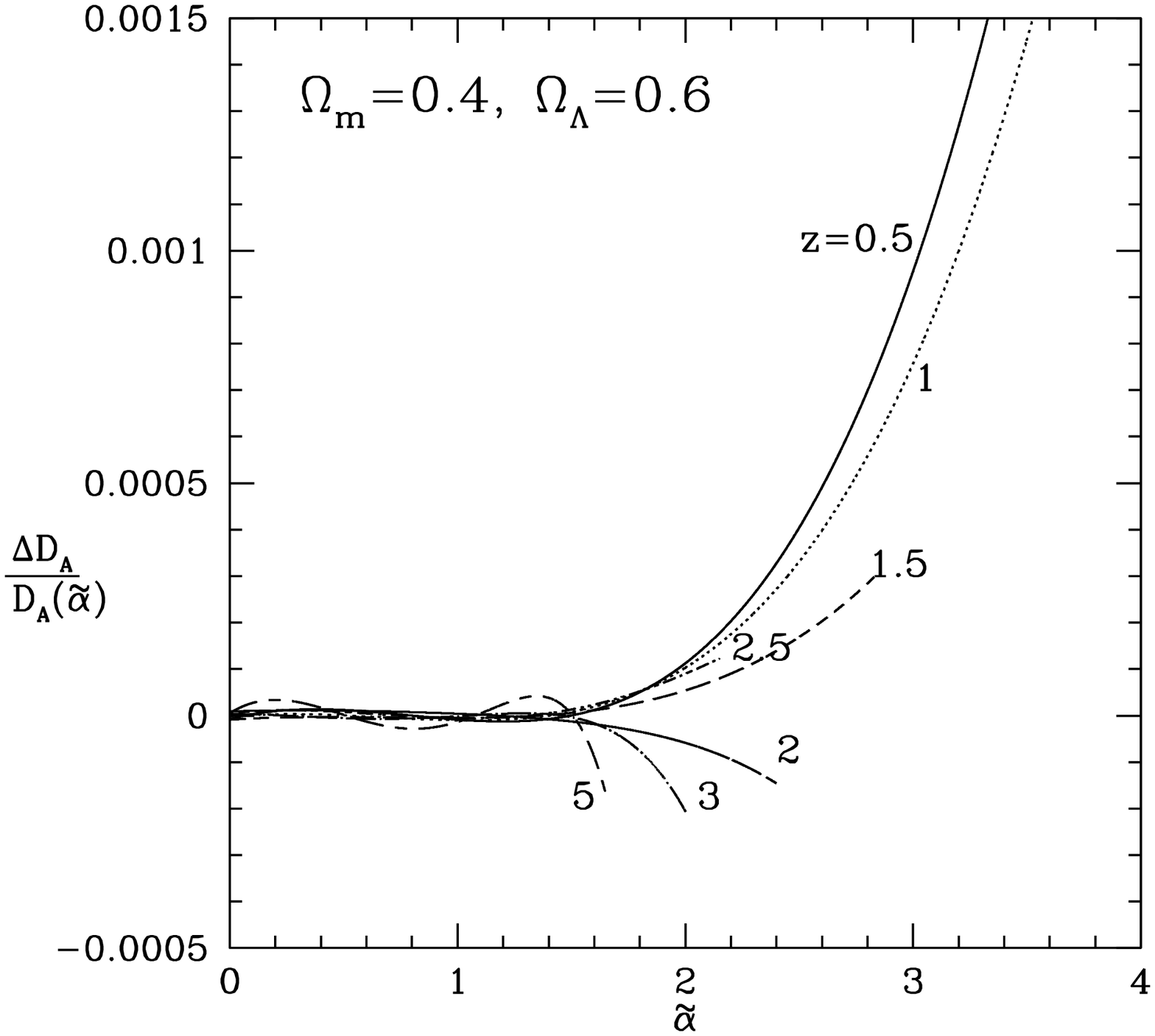]{
The relative differences between the exact angular diameter distance and the
analytical approximation given by Eq.(\ref{eq:DRa}) as function of the
smoothness parameter $\tilde{\alpha}$, for z=0.5 (solid line),
1 (dotted line), 1.5 (short dashed line), 2 (long dashed line), 2.5 
(dot - short dashed line), 3 (dot - long dashed line), and 5 (short dash -
long dashed line).
Note that $\tilde{\alpha}<1$ in underdense beams, while 
$\tilde{\alpha}>1$ in overdense beams.}

\figcaption[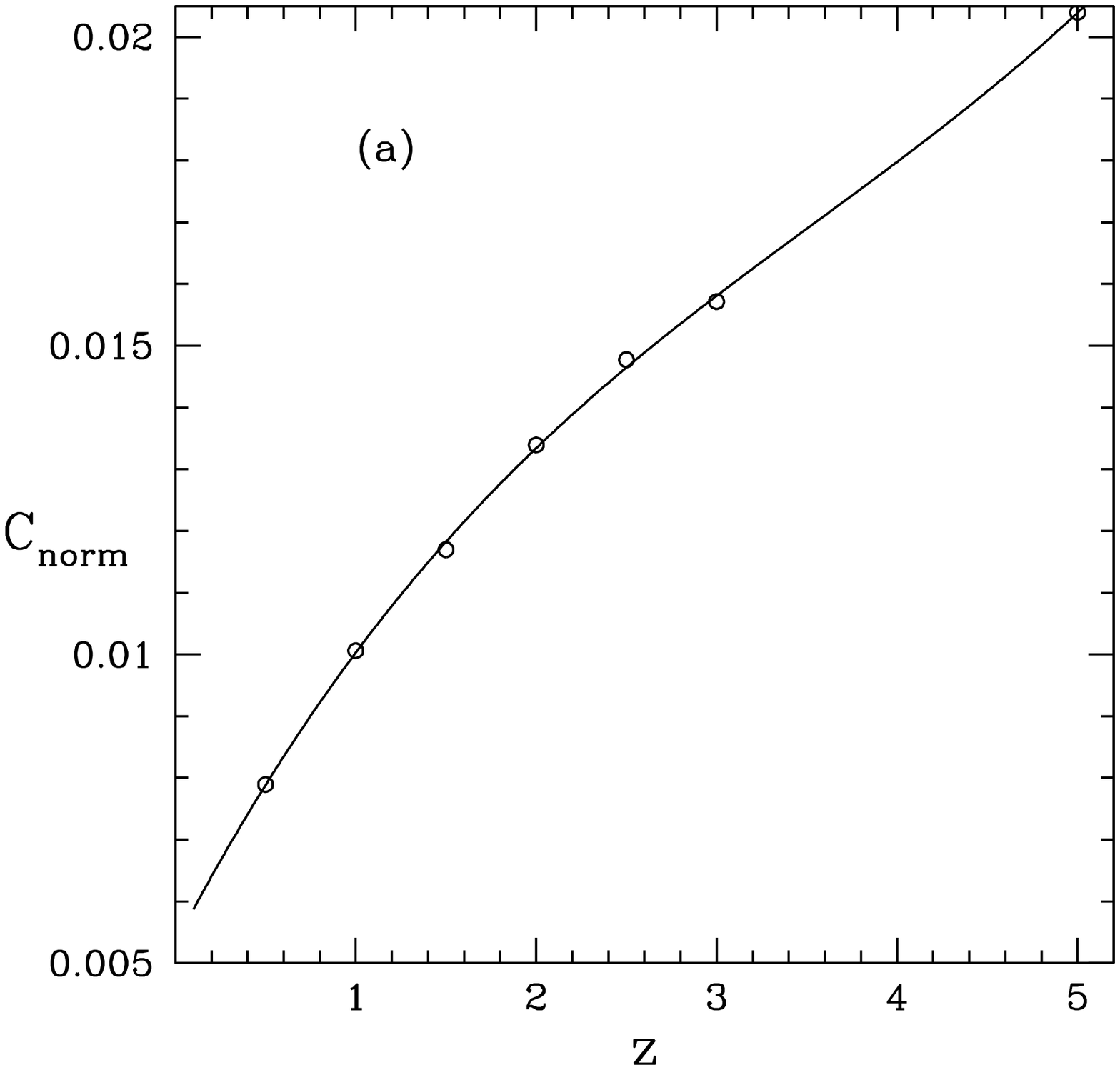]{
The coefficients $C_{norm}$, $\tilde{\alpha}_{peak}$, $w$, and $q$ as functions
of $z$; the points are numerical results, the solid curves are analytical
fits given by Eq.(\ref{eq:aq}).}

\figcaption[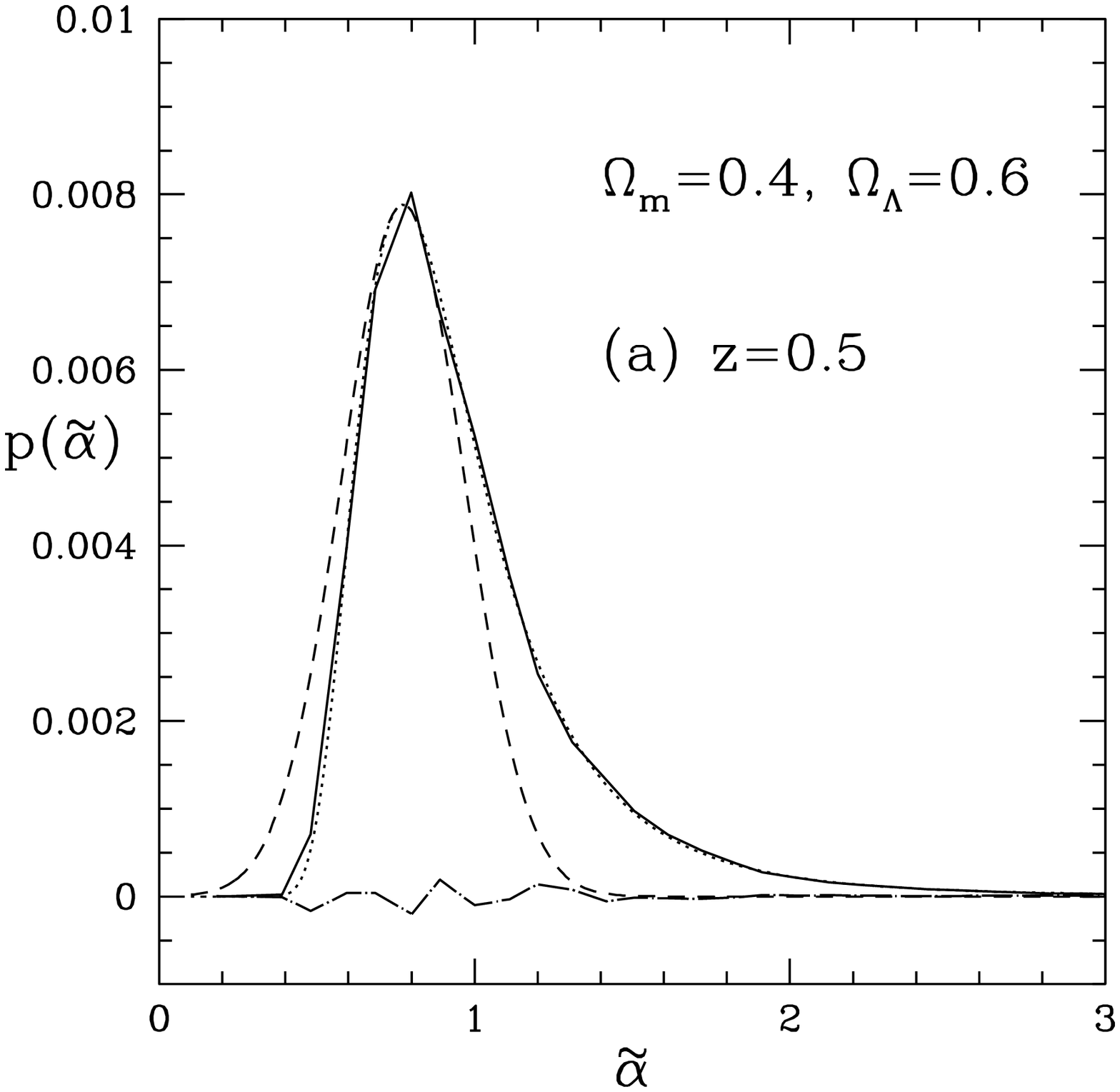]{
The distribution of the direction dependent smoothness parameter,
$p(\tilde{\alpha}|z)$, for (a) $z=0.5$, (b) $z=2$, and 
(c) $z=5$. The solid line
is derived from $p(\mu|z)$ found numerically by Wambsganss et al. (1997);
the dotted line is given by Eq.(\ref{eq:p(alpha)}), with coefficients
from Eq.(\ref{eq:aq}); the dot-dash line shows the difference between
the solid curve and the dotted curve;
the dashed line shows the Gaussian distribution
given by Eq.(\ref{eq:pGauss}).
It is difficult to see the dotted lines because they are so close 
to the solid lines.
Note that $\tilde{\alpha}<1$ in underdense beams, while 
$\tilde{\alpha}>1$ in overdense beams.}

\figcaption[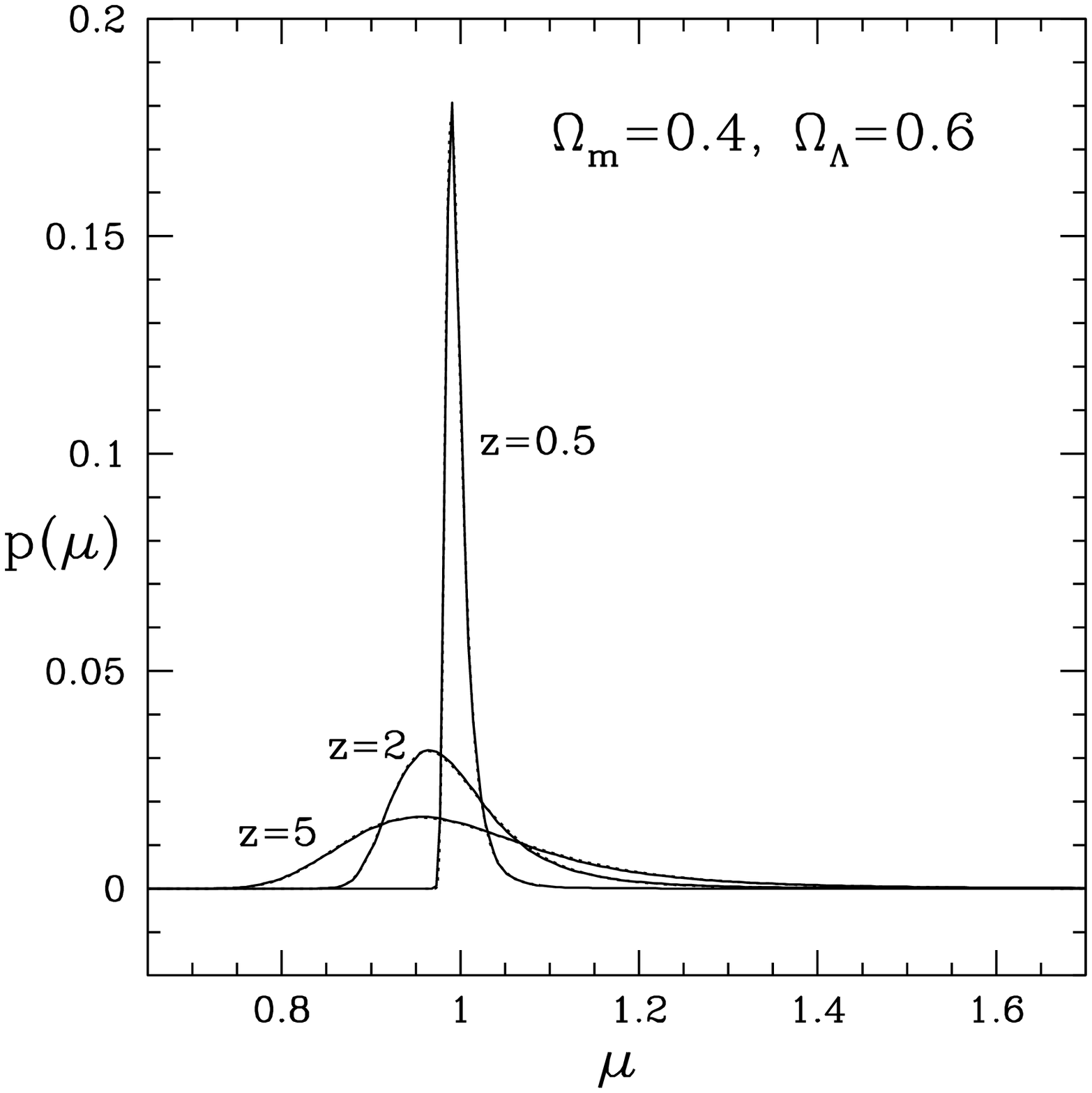]{
The magnification distribution of standard candles, $p(\mu|z)$, 
for $z=0.5$, 2, and 5. The solid line
is the $p(\mu|z)$ found numerically by Wambsganss et al. (1997);
the dotted line is given by Eq.(\ref{eq:p(mu)}), with $p(\tilde{\alpha}|z)$
given by Eq.(\ref{eq:p(alpha)}).
It is difficult to see the dotted lines because they are so close 
to the solid lines.
}

\figcaption[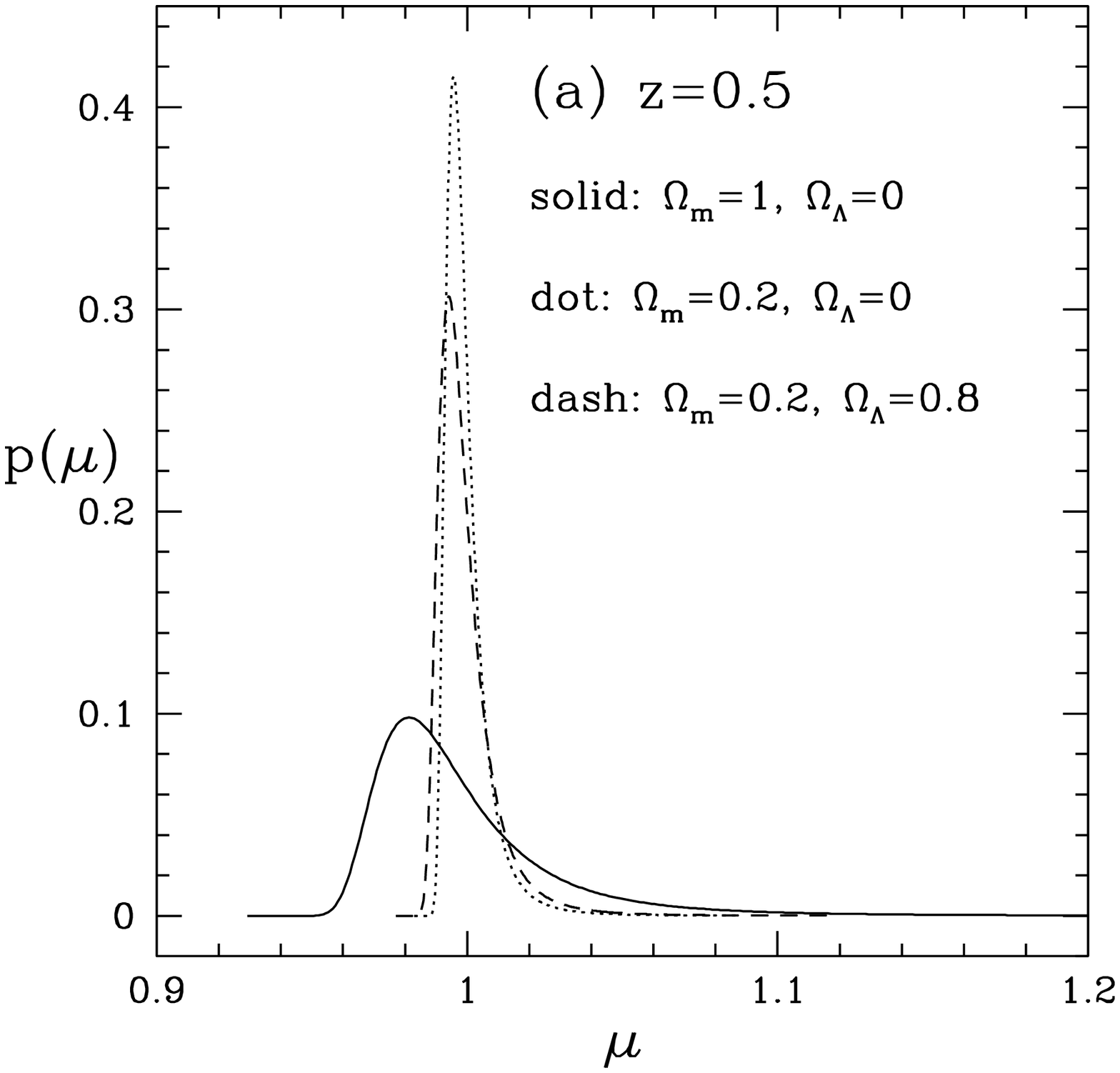]{
$p(\mu|z)$ at $z=0.5$ (a), 2 (b), and 5 (c) for three 
cosmological models:
(1) $\Omega_m=1$ and $\Omega_{\Lambda}=0$;
(2) $\Omega_m=0.2$ and $\Omega_{\Lambda}=0$;
(3) and $\Omega_m=0.2$ and $\Omega_{\Lambda}=0.8$.
Note that Fig.5 and Fig.4 have the same matter distribution
(the same $p(\tilde{\alpha}|z)$), but different cosmological parameters.}

\clearpage

\setcounter{figure}{0}
\plotone{f1.eps}
\figcaption[f1.eps]{
The relative differences between the exact angular diameter distance and the
analytical approximation given by Eq.(\ref{eq:DRa}) as function of the
smoothness parameter $\tilde{\alpha}$, for z=0.5 (solid line),
1 (dotted line), 1.5 (short dashed line), 2 (long dashed line), 2.5 
(dot - short dashed line), 3 (dot - long dashed line), and 5 (short dash -
long dashed line). Note that $\tilde{\alpha}<1$ in underdense beams, while 
$\tilde{\alpha}>1$ in overdense beams.}

\plotone{f2a.eps}
\figcaption[f2a.eps]{
The coefficients $C_{norm}$, $\tilde{\alpha}_{peak}$, $w$, and $q$ as functions
of $z$; the points are numerical results, the solid curves are analytical
fits given by Eq.(\ref{eq:aq}).
(a) $C_{norm}$.}
\setcounter{figure}{1}
\plotone{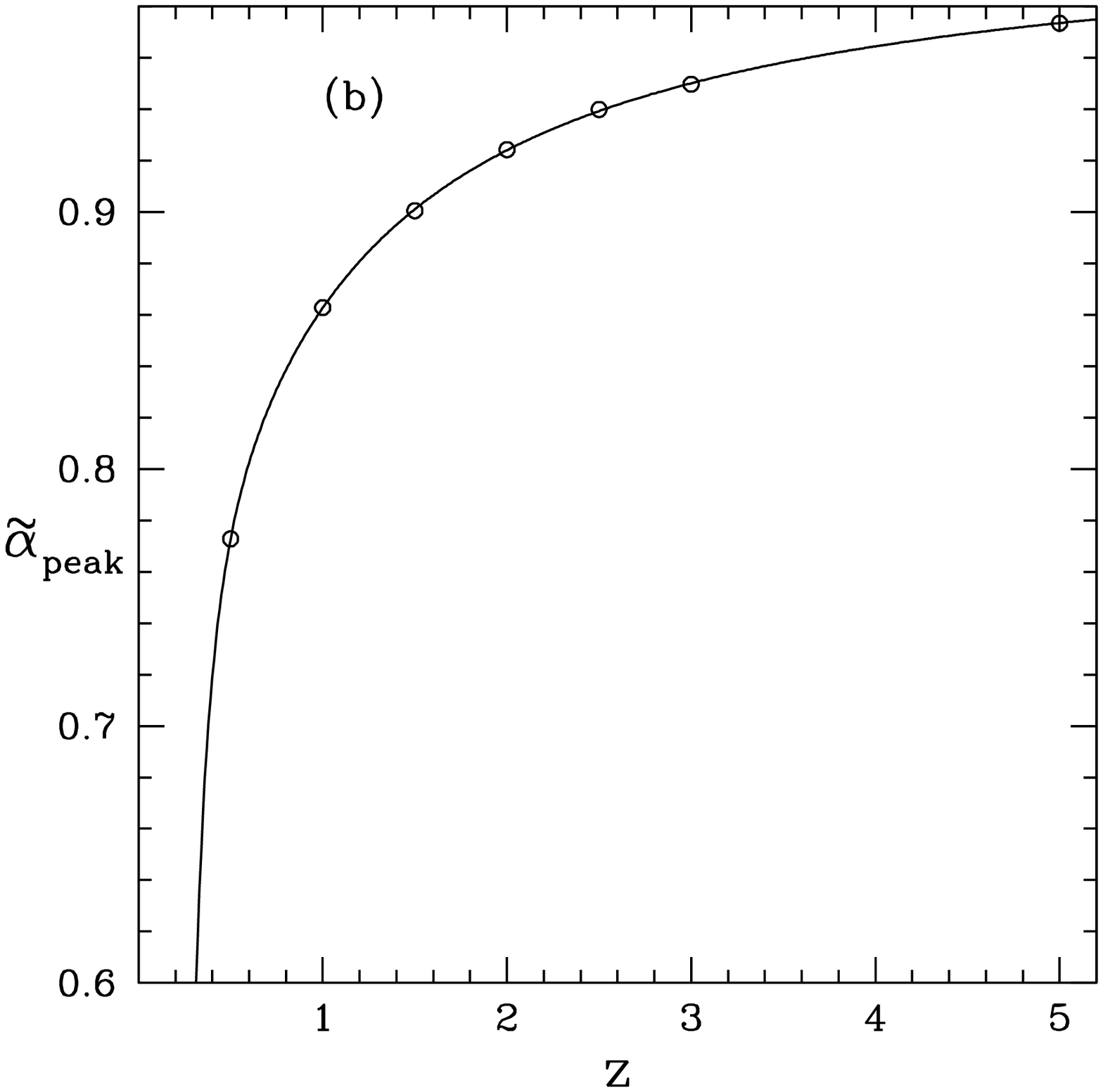}
\figcaption[f2b.eps]{ (b) $\tilde{\alpha}_{peak}$.}
\setcounter{figure}{1}
\plotone{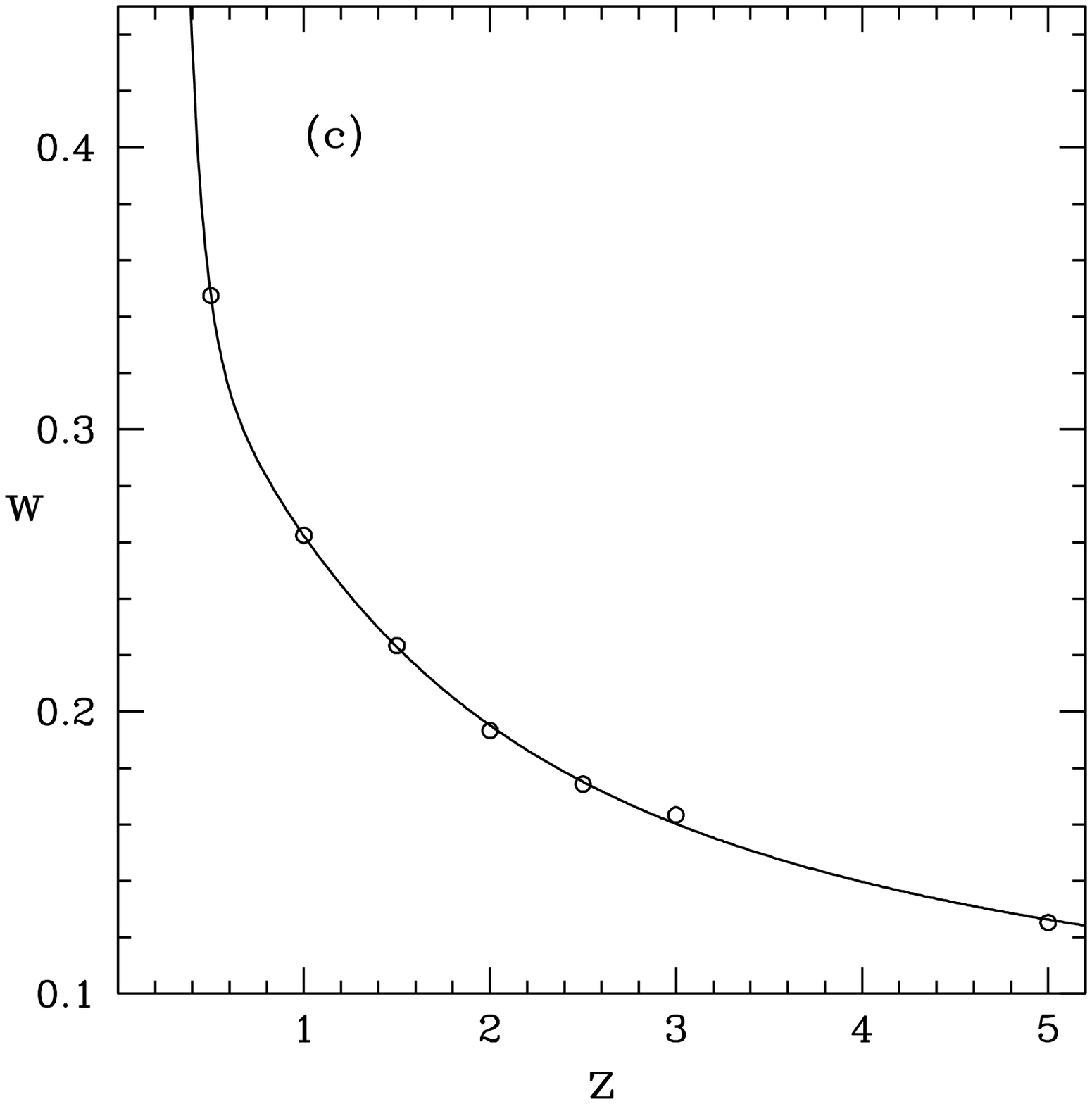}
\figcaption[f2c.eps]{ (c) $w$.}
\setcounter{figure}{1}
\plotone{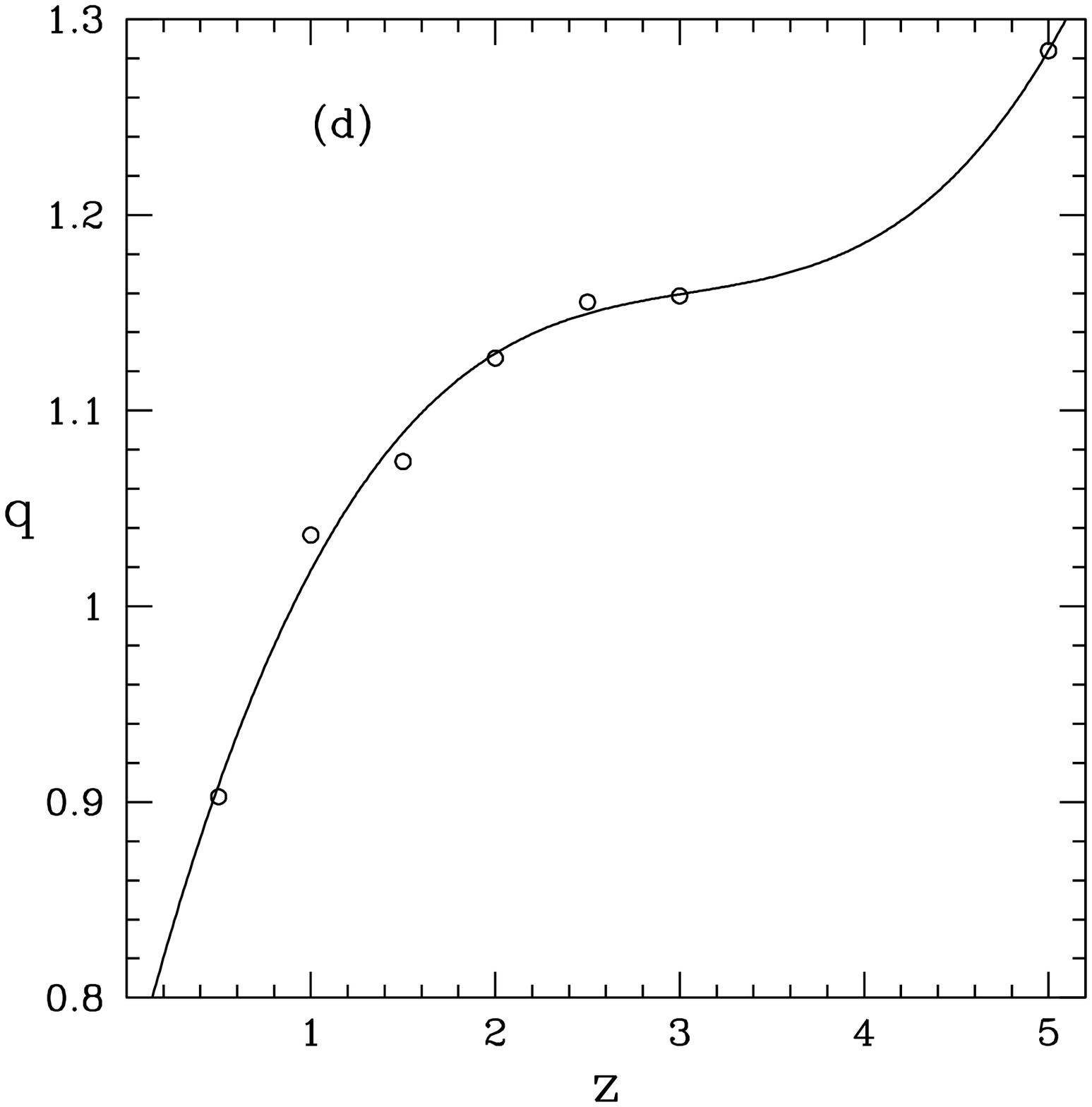}
\figcaption[f2d.eps]{ (d) $q$.}

\plotone{f3a.eps}
\figcaption[f3a.eps]{
The distribution of the direction dependent smoothness parameter,
$p(\tilde{\alpha}|z)$. The solid line
is derived from $p(\mu|z)$ found numerically by Wambsganss et al. (1997);
the dotted line is given by Eq.(\ref{eq:p(alpha)}), with coefficients
from Eq.(\ref{eq:aq}); the dot-dash line shows the difference between
the solid curve and the dotted curve;
the dashed line shows the Gaussian distribution
given by Eq.(\ref{eq:pGauss}).
It is difficult to see the dotted lines because they are so close 
to the solid lines.
Note that $\tilde{\alpha}<1$ in underdense beams, while 
$\tilde{\alpha}>1$ in overdense beams.
(a) $z=0.5$.}
\setcounter{figure}{2}
\plotone{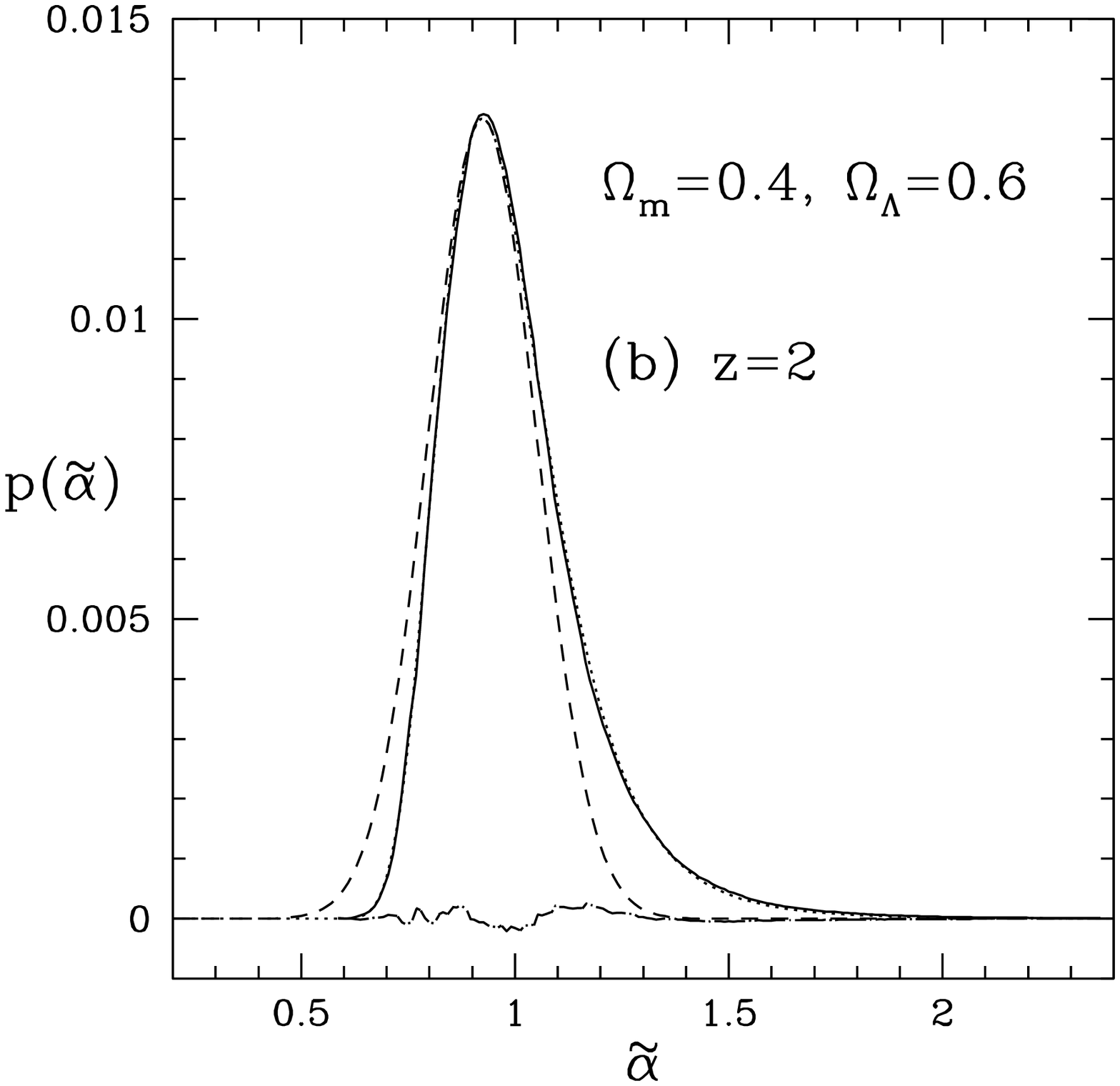}
\figcaption[f3b.eps]{(b) $z=2$.}
\setcounter{figure}{2}
\plotone{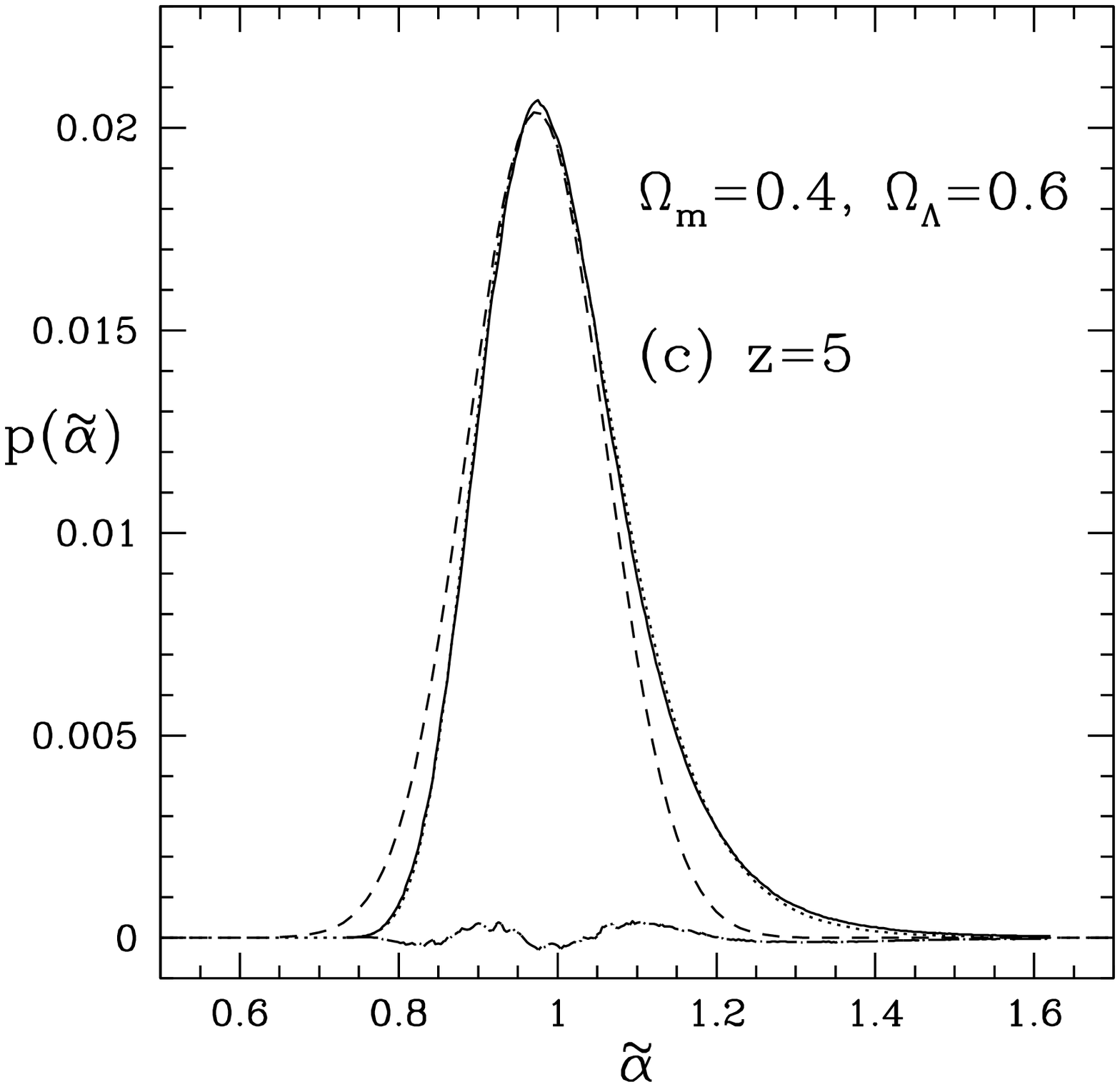}
\figcaption[f3c.eps]{(c) $z=5$.}

\plotone{f4.eps}
\figcaption[f4.eps]{
The magnification distribution of standard candles, $p(\mu|z)$, 
for $z=0.5$, 2, and 5. The solid line
is the $p(\mu|z)$ found numerically by Wambsganss et al. (1997);
the dotted line is given by Eq.(\ref{eq:p(mu)}), with $p(\tilde{\alpha}|z)$
given by Eq.(\ref{eq:p(alpha)}).
It is difficult to see the dotted lines because they are so close 
to the solid lines.}

\plotone{f5a.eps}
\figcaption[f5a.eps]{
$p(\mu|z)$ for three cosmological models:
(1) $\Omega_m=1$ and $\Omega_{\Lambda}=0$;
(2) $\Omega_m=0.2$ and $\Omega_{\Lambda}=0$;
(3) and $\Omega_m=0.2$ and $\Omega_{\Lambda}=0.8$.
Note that Fig.5 and Fig.4 have the same matter distribution
(the same $p(\tilde{\alpha})$), but different cosmological parameters.
(a) $z=0.5$.}
\setcounter{figure}{4}
\plotone{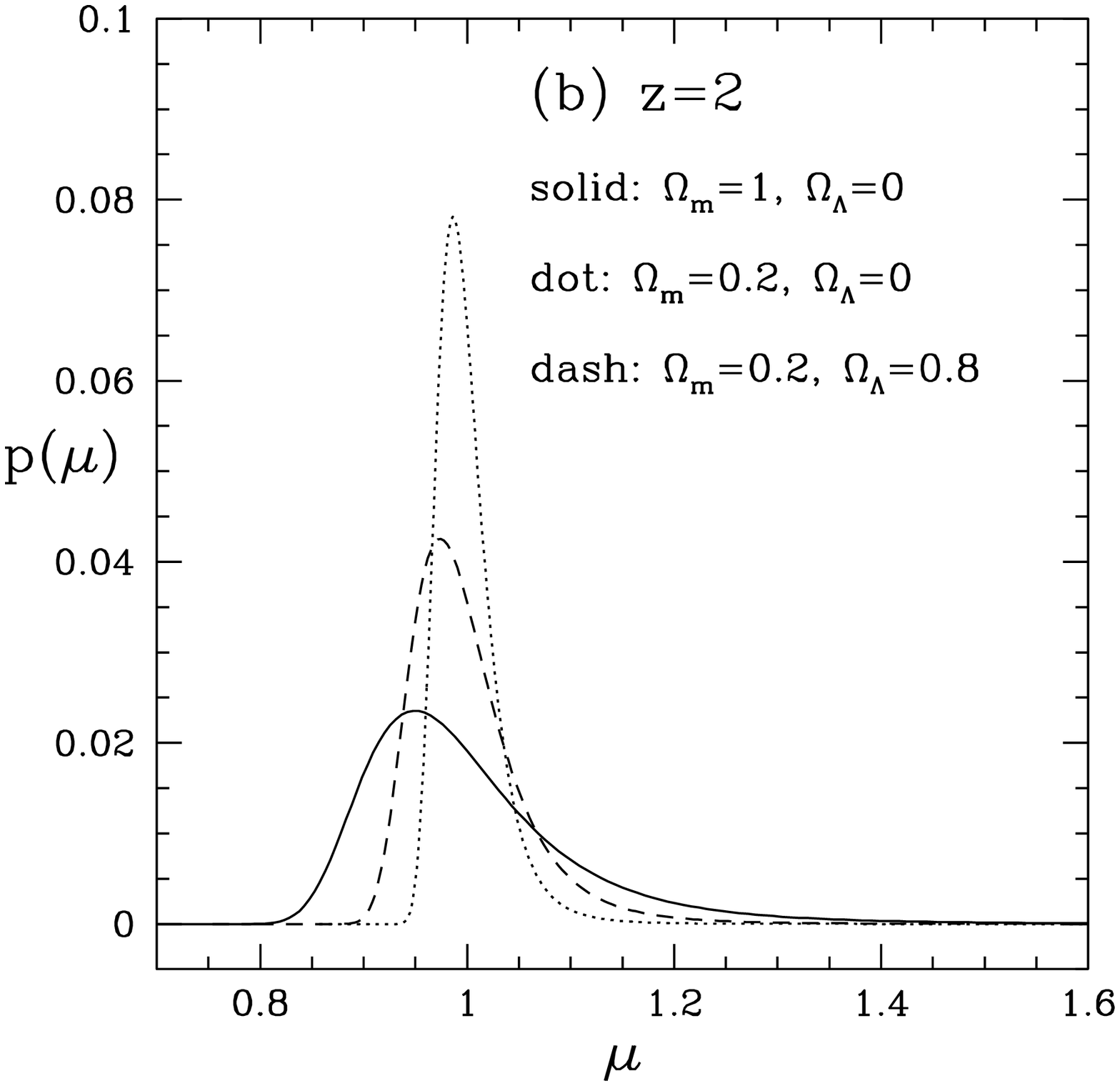}
\figcaption[f5b.eps]{ (b) $z=2$.}
\setcounter{figure}{4}
\plotone{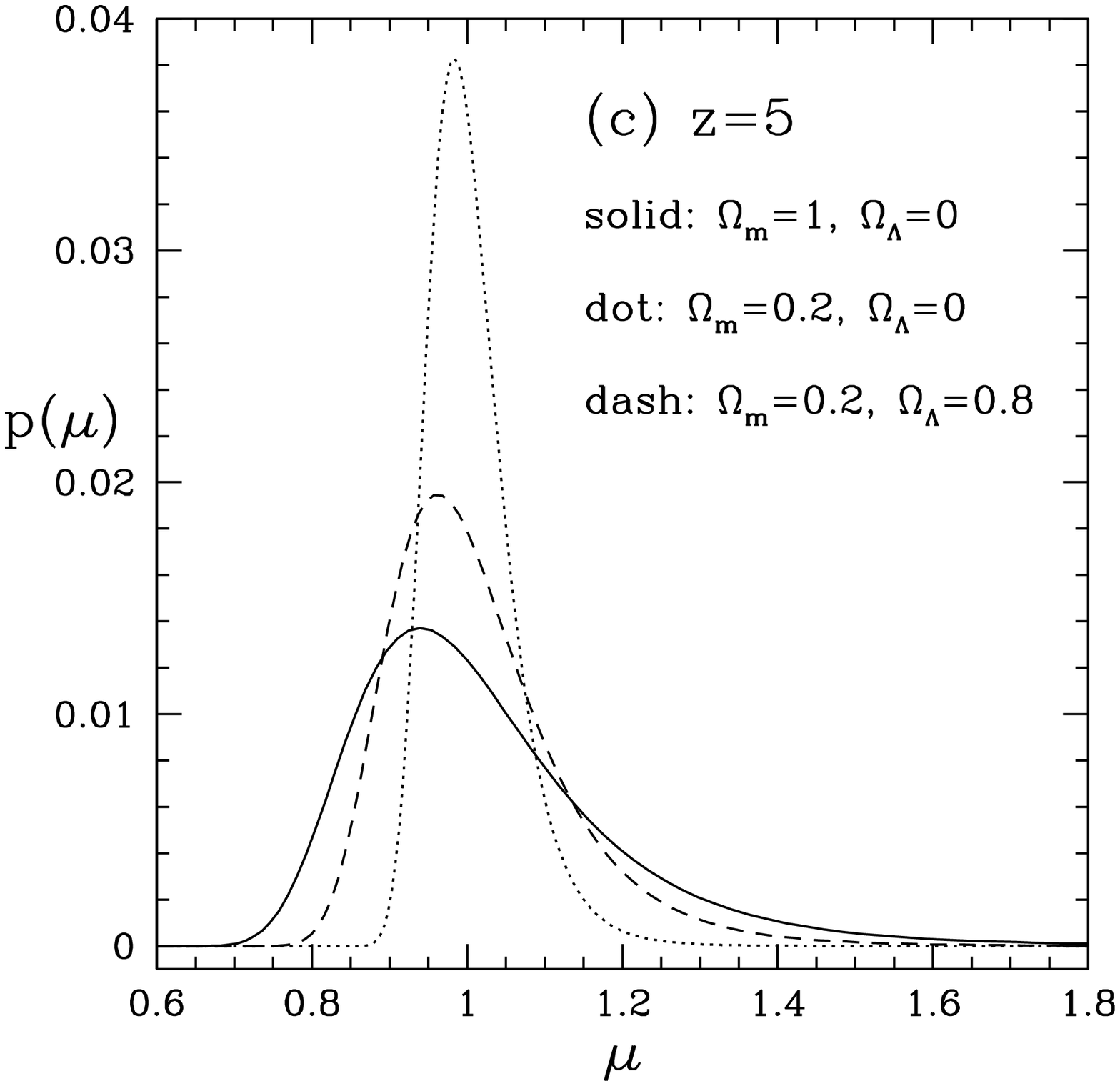}
\figcaption[f5c.eps]{ (c) $z=5$.}

\end{document}